\documentclass[aps,prl,showpacs,twocolumn,superscriptaddress,nofootinbib,groupedaddress]{revtex4} 
\usepackage{color}
\usepackage{graphicx}
\usepackage{amsmath}
\usepackage{amssymb}
\usepackage[colorlinks=true,citecolor=darkred,urlcolor=darkred, pdfborder={0 0 0}]{hyperref}
\usepackage[makeroom]{cancel}
\usepackage[normalem]{ulem}
\usepackage{soul}
\definecolor{darkred}{rgb}{0.6,0,0}

\definecolor{linkcolor}{rgb}{0,0,0.5}

\def\gsim{\raise0.3ex\hbox{$\;>$\kern-0.75em\raise-1.1ex\hbox{$\sim\;$}}}
\def\lsim{\raise0.3ex\hbox{$\;<$\kern-0.75em\raise-1.1ex\hbox{$\sim\;$}}}

\def\beqn#1{\begin{equation}\label{#1}}
\def\eeqn{\end{equation}}

\def\beqa#1{\begin{eqnarray}\label{#1}}
\def\eeqa{\end{eqnarray}}

\def\Z2{$\mathcal{Z_2}$}

\def\vev#1{\left\langle #1\right\rangle}

\newcommand {\ignore}[1]{}

\begin{document}

%\author{Mario Reig}\email{mario.reig@ific.uv.es}
%\affiliation{\AddrAHEP}

\title{Cosmic implications of a low-scale solution to the axion domain wall problem}
%\title{On the Low-scale Instanton Interference Solution to the Axion Domain Wall Problem}

%%%%%%%%%%%%%%%%%%%%%%%%%%%%%%%%%%%%%%%%%%%%%%%%%%%%%%%%%

%\begin{textblock*}{5cm}(11cm,-8.2cm)
%\end{textblock*}

%%%%%%%%%%%%%%%%%%%%%%%%%%%%%%%%%%%%%%%%%%%%%%%%%%
\author{Andrea Caputo}\email{andrea.caputo@uv.es}\affiliation{Instituto de Fisica Corpuscular, Universidad de Valencia and CSIC, Edificio Institutos Investigacion, Catedratico Jose Beltran 2, Paterna, 46980 Spain}
\author{Mario Reig}\email{mario.reig@ific.uv.es}\affiliation{Instituto de Fisica Corpuscular, Universidad de Valencia and CSIC, Edificio Institutos Investigacion, Catedratico Jose Beltran 2, Paterna, 46980 Spain}

\begin{abstract} 
The post-inflationary breaking of Peccei-Quinn (PQ) symmetry can lead to the cosmic domain wall catastrophe. In this paper we show how to avoid domain walls by implementing the instanton interference effect with a new interaction which itself breaks PQ symmetry and confines at an energy scale smaller than $\Lambda_{\text{QCD}}$. We give a general description of the mechanism and consider its cosmological implications and constraints within a minimal model. Contrary to other mechanisms we do not require an inverse phase transition or fine-tuned bias terms. Incidentally, the mechanism leads to the introduction of new self-interacting dark matter candidates and the possibility of producing gravitational waves in the frequency range of SKA. Unless a fine-tuned hidden sector is introduced, the mechanism predicts a QCD axion in the mass range $1-15\text{ meV}$. 
\end{abstract}

\maketitle

\medskip

%%%%%%%%%%%%%%%%%%%%%%%%%%%%%%%%%%%%%%%%%%%%%%%%%%%%%%%%%

\section{Introduction and motivation}
\label{sec:intro}
%%%%%%%%%%%%%%%%%%%%%%%%%%%%%%%%%%%%%%%%%%%%%%%%%%%%%%%%%
The axion solution to the strong $CP$ problem \cite{Peccei:1977hh,Wilczek:1977pj,Weinberg:1977ma, Dine:1981rt,Zhitnitsky:1980tq,Kim:1979if} is a well-known paradigm where domain walls bounded by strings\footnote{This kind of topological defect was first proposed in Ref.  \cite{Kibble:1982dd} and has been experimentally found in superfluid $^3$He \cite{2019NatCo..10..237M}. } emerge \cite{Vilenkin:1982ks,Sikivie:1982qv}. This is because as the Universe cools down two different phase transitions occur. In the first one the Peccei-Quinn (PQ) scalar 
\begin{equation}
\Phi=(v+\rho)e^{ia/v}/\sqrt{2}\,,
\end{equation}
develops a nonzero vacuum expectation value (VEV), $|\langle\Phi\rangle|=v/\sqrt{2}$, breaking $U(1)_{PQ}$ spontaneously, and cosmic strings form. This is expected to happen at very high temperatures, around $T\sim v$. When the cosmic temperature reaches the QCD scale, QCD instantons generate an effective potential for the axion field and domain walls form. At this point, cosmic strings become attached to domain walls. The potential will have a periodicity determined by $N_{\text{QCD}}$, the QCD coefficient anomaly, computed as
\begin{equation}
N_{\text{QCD}}=2\sum q_Rt_R\,,
\label{anomaly}
\end{equation}
where $t_{R}$ is the Dynkin index and  $q_R$ is the PQ charge of the fermions. Notice that the axion decay constant and the VEV, $|\langle\Phi\rangle|=v$, are related as
	\begin{equation}
	f_a=\frac{v}{N_{\text{QCD}}}\,.
	\end{equation}
It is well known that for $N_{\text{QCD}}=1$ the string-wall system is not stable because the tension of the walls make strings collapse to a point and the network quickly decays into nonrelativistic axions \cite{Vilenkin:1981zs}. However, for $N_{\text{QCD}}>1$ each string gets attached to $N_{\text{QCD}}$ walls and the network cannot decay: the walls are topologically protected and stable. Their evolution with cosmic expansion is slower than that of matter or radiation and will eventually dominate the energy density of the Universe. To avoid such a cosmological catastrophe one needs to make them disappear or prevent their formation. Actually, as shown in Ref. \cite{DiLuzio:2016sbl}, simple Kim-Shifman-Vainshtein-Zhakharov (KSVZ) axion models with different representations for the exotic quarks generating the PQ anomaly generically have  $N_{\text{QCD}} > 1$  and, therefore, do suffer from the DW problem. Therefore, simply neglecting the problem and assume $N_{DW}=1$ sounds too simplistic.

A straightforward way to solve the domain wall problem is to evoke cosmic inflation \cite{Guth:1980zm,Linde:1981mu,Albrecht:1982wi}. Domain walls are, in this case, pushed beyond the horizon and will not harm our cosmic evolution. This is the case of a PQ symmetry spontaneously broken before inflation and never restored after reheating. This case, however, can be constrained by isocurvature fluctuations in the cosmic microwave background  \cite{Hertzberg:2008wr}. \footnote{Isocurvature fluctuations might be suppressed if Peccei-Quinn symmetry was explicitly broken by a non-Abelian interaction in the early Universe, as proposed in \cite{Takahashi:2015waa}. This scenario could be tested in the near future since it predicts self-interacting dark radiation with $\Delta N_{\text{eff}}=O(0.01-0.1)$ .}

Here instead we consider a scenario in which PQ symmetry is broken after inflation. Different post-inflation mechanisms have been proposed since the 1980s. In the Lazarides-Shafi mechanism \cite{Lazarides:1982tw}, for example, one associates the discrete symmetry unbroken by instantons to the center of a gauge symmetry\footnote{See Ref. \cite{Barr:1982bb} for a different realization in the context of family symmetries.}. Other solutions such as primordial black hole formation \cite{Stojkovic:2005zh} or the Witten effect have been explored more recently \cite{Kawasaki:2015lpf,Sato:2018nqy}.

As a guideline, to make the walls unstable one needs to remove their topological protection by explicitly breaking the discrete symmetry that relates the degenerated set of vacua. The simplest solution that follows this philosophy is almost as old as the DW problem and is known as the bias term solution \cite{Sikivie:1982qv}. It consists in adding to the scalar potential an \textit{ad hoc} term
\begin{equation}
\Delta V_{bias}=\Xi f_a^3\left(\Phi e^{i\delta}+\text{H.c.}\right)\,.
\end{equation}
This term explicitly breaks PQ symmetry and produces a potential for the axion field that generates an effective theta term which is constrained to be $\theta\leq 10^{-10}$. Then, if one assumes \textit{natural} values for the phase, $\delta\sim O(1)$, the dimensionless parameter $\Xi$ is constrained to be \cite{Kawasaki:2014sqa}
\begin{equation}
\Xi\leq 2\cdot 10^{-45}N_{DW}^{-2}\Big(\frac{10^{10}\text{ GeV}}{f_a}\Big)^4 \,.
\end{equation} 

In this paper we will consider a natural (and not fine-tuned) realization of the bias term by using a low-scale version of the instanton interference effect (IIE). The introduction of a new confining interaction and its associated instantons will generate the explicit breaking of the $Z_{N_{\text{DW}}}$ symmetry, unbroken by QCD instantons. 

The paper is organized as follows. We first introduce the IIE and explain how it solves the domain wall problem. Then, we explore the constraints on the confinement scale of the HC sector we introduce. Once we set the scale of the new sector, we specify to a minimal $SU(N)$ model and explore its cosmological implications. Finally, we conclude and comment on future directions to follow.

\section{The Instanton Interference Effect}
The IIE is a compelling mechanism to avoid the cosmic domain wall problem in an inflation-independent way (it applies in both, pre- and post-inflationary PQ-breaking scenarios) \cite{Barr:2014vva,Reig:2019vqh}. In this mechanism one adds to the invisible QCD axion model a new non-Abelian gauge group $HC$ which is also anomalous under PQ symmetry,
\begin{equation}
HC\times U(1)_{PQ}\times \text{SM}\,,
\end{equation}
where SM is the usual Standard Model gauge symmetry $SU(3)_C\times SU(2)_L\times U(1)_Y$. Instantons associated with the new group will in general break the PQ symmetry down to a $Z_{N_{HC}}$ subgroup. On the other hand, QCD instantons break the same symmetry to a $Z_{N_{\text{QCD}}}$ discrete symmetry. The full axion potential reads
\begin{equation}\label{axion_potential}\begin{split}
V(a)=&V_{QCD}(a)+V_{HC}(a)=\kappa\Lambda_{QCD}^4\left(1-\cos \left(\frac{a}{v}N_{QCD}\right)\right)\\&+\Lambda_{HC}^4\left(1-\cos \left(\frac{a}{v}N_{HC}-\delta\right)\right)\,,
\end{split}
\end{equation}
with $\kappa\approx 1.2\times 10^{-3}$, in which we recognize the QCD term with anomaly coefficient $N_{\text{QCD}}$, and the new term due to the $HC$ interaction which confines at some scale $\Lambda_{HC}$ and has its own anomaly coefficient, $N_{HC}$. When $N_{\text{QCD}}$ and $N_{HC}$ are coprime numbers, the PQ symmetry is completely broken by the combination of explicit symmetry breaking by instantons and the domain wall problem is solved. This is the IIE. We notice that a similar periodic bias term with $N_{HC}=1$ has been considered in \cite{Ferrer:2018uiu}; there, the authors focused on the generation of primordial black holes and did not consider its origin or its cosmological implications (and constraints), which are intriguing and diverse. 

In its high-scale realization \cite{Barr:2014vva,Reig:2019vqh}, the IIE mechanism requires a confinement scale much larger than the QCD scale. In fact, if PQ symmetry is spontaneously broken by hypercolor HC condensates, as in Ref. \cite{Reig:2019vqh}, the $f_a$ and $\Lambda_{HC}$ scales coincide. Therefore one needs to turn off the interaction below a critical temperature, once PQ is spontaneously broken and the axion field sits on the same vacuum everywhere, in order to allow the standard PQ mechanism to work and dynamically drive the axion field to a $CP$-conserving minimum. To allow this, a new fermion in the HC sector and a new scalar with an inverted phase transition are needed. 

It is attractive to explore the possibility of a low-scale version of the IIE with $\Lambda_{HC}\ll \Lambda_{\text{QCD}}$, which is indeed the goal of the present work. As we will show, this scenario does not require any of the complications related to the inverse phase transition and can have peculiar cosmological implications, such as the presence of new self-interacting dark matter candidates and a strong first-order phase transition which could produce detectable gravitational waves.

Finally, we also note that the bias term introduced in the PQ potential might also be generated by gravitational effects \cite{Rai:1992xw}. However the standard lore is that gravity effects arise at the nonperturbative level \cite{Kallosh:1995hi, Alonso:2017avz} with a suppression proportional to $\propto e^{-M_{Pl}/f_a}$. For the range of values considered in the present work these effects are then negligible.

\section{Setting the confinement scale of the dark sector}
In this section we derive upper and lower bounds to the HC scale, $\Lambda_{HC}$. To this end, we do not want to spoil the solution to the strong CP problem or overclose the Universe, which would produce too much axion dark matter through domain wall decay.

\subsection{Lower bounds to $\Lambda_{HC}$}
Two different lower bounds appear for the confinement scale $\Lambda_{HC}$ of the new interaction. When the walls decay to DM, as it is our QCD axion, the big bang nucleosynthesis (BBN) constraints are relaxed and the only requirement is that they must disappear before matter-radiation equality, given roughly by $T_{\text{eq}}\sim 1$ eV. The reason is that since these domain walls can only decay into axions or gravitational waves (gravitons) the interaction of their decay products with the SM is not efficient and they do not spoil BBN. In general, for $N_{\text{DW}}>1$, domain walls will disappear when the discrete symmetry $Z_{N_{\text{QCD}}}$ unbroken by QCD instantons starts to \textit{feel} the effects of HC instantons. Then, due to the IIE the $Z_{N_{\text{QCD}}}$ symmetry gets explicitly broken and the domain walls are not topologically protected. This occurs because the initially degenerated vacua get a small splitting from the bias term. Therefore, the new term in the potential causes an energy difference between the false and true minima $\Delta\rho \sim \Lambda_{HC}^4$, leading to a pressure $p_V \sim \Delta\rho$ which acts against the domain wall. We remark that, in our case, the bias term comes from the potential $V_{HC}(a)$ (generated by HC instantons). The condition that the vacuum pressure generated from the energy difference exceeds the wall tension reads
\begin{equation}
	p_V > p_{tension}\,,
\end{equation}
and therefore
\begin{equation}
V_{bias}>H_{decay}\sigma\,,
\end{equation}
with $\sigma=8m_af_a^2$. This allows us to write the Hubble parameter when the walls decay as:
\begin{equation}\label{Hdecay}
H_{decay}=\frac{V_{\text{bias}}}{\sigma}\approx\frac{\Lambda_{HC}^4}{8m_af_a^2}\,.
\end{equation}
We can impose that the domain walls decay before matter domination and thus obtain the lower bound to the confinement scale,
\begin{equation}\label{constrain_matter_rad}
\Lambda_{HC}\gg 3.36\times 10^{-10}\left[8\Lambda_{\text{QCD}}^2f_a \right]^{1/4}\text{GeV}^{1/4}\,.
\end{equation}
A different (and actually more stringent) bound can be obtained by the requirement that DWs never dominate the energy density of the Universe. The contribution of domain walls to the energy density of the Universe is 
\begin{equation}
	\rho_w \sim \frac{\sigma t^2}{t^3} \sim \frac{\sigma}{t}\,.
\end{equation}
The Universe becomes dominated by domain walls  when they reach relativistic speeds at the time $t_c \sim (G\sigma)^{-1}$. Consequently, one has to impose that the time associated with the decay is much shorter:
\begin{equation}
t_c\sim \frac{1}{G\sigma}\gg H_{decay}^{-1}\,.
\end{equation}
Using Eq. (\ref{Hdecay}), we get
\begin{equation}
\Lambda_{HC}\gg %0.1\left(\frac{8}{N^2}\right)^{1/2}%
0.3\left(\frac{f_a}{M_P}\right)^{1/2}\Lambda_{\text{QCD}}\,,
\end{equation}
which looks more stringent than Eq.(\ref{constrain_matter_rad}) for reasonable values of $f_a$.

\subsection{Upper bound to $\Lambda_{HC}$}
The axion potential [see Eq. (\ref{axion_potential})] gives rise to two contributions from QCD and HC instantons. The QCD contribution arises first. Then, below a critical temperature, the HC potential also arises and a small $\theta_{\text{eff}}$ is generated. We can minimize the above potential to get
\begin{equation}
\frac{\partial V}{\partial a}=0\rightarrow \frac{\vev{a}}{v}\approx \frac{-\Lambda_{HC}^4\sin \delta}{\kappa\Lambda_{\text{QCD}}^4N_{\text{QCD}}+\Lambda_{HC}^4N_{HC}\cos\delta}\,,
\end{equation}
and impose that the solution of the strong \textit{CP} problem is not spoiled. 

\begin{figure}[!htb!]
\centering
\includegraphics[width=1.05\linewidth]{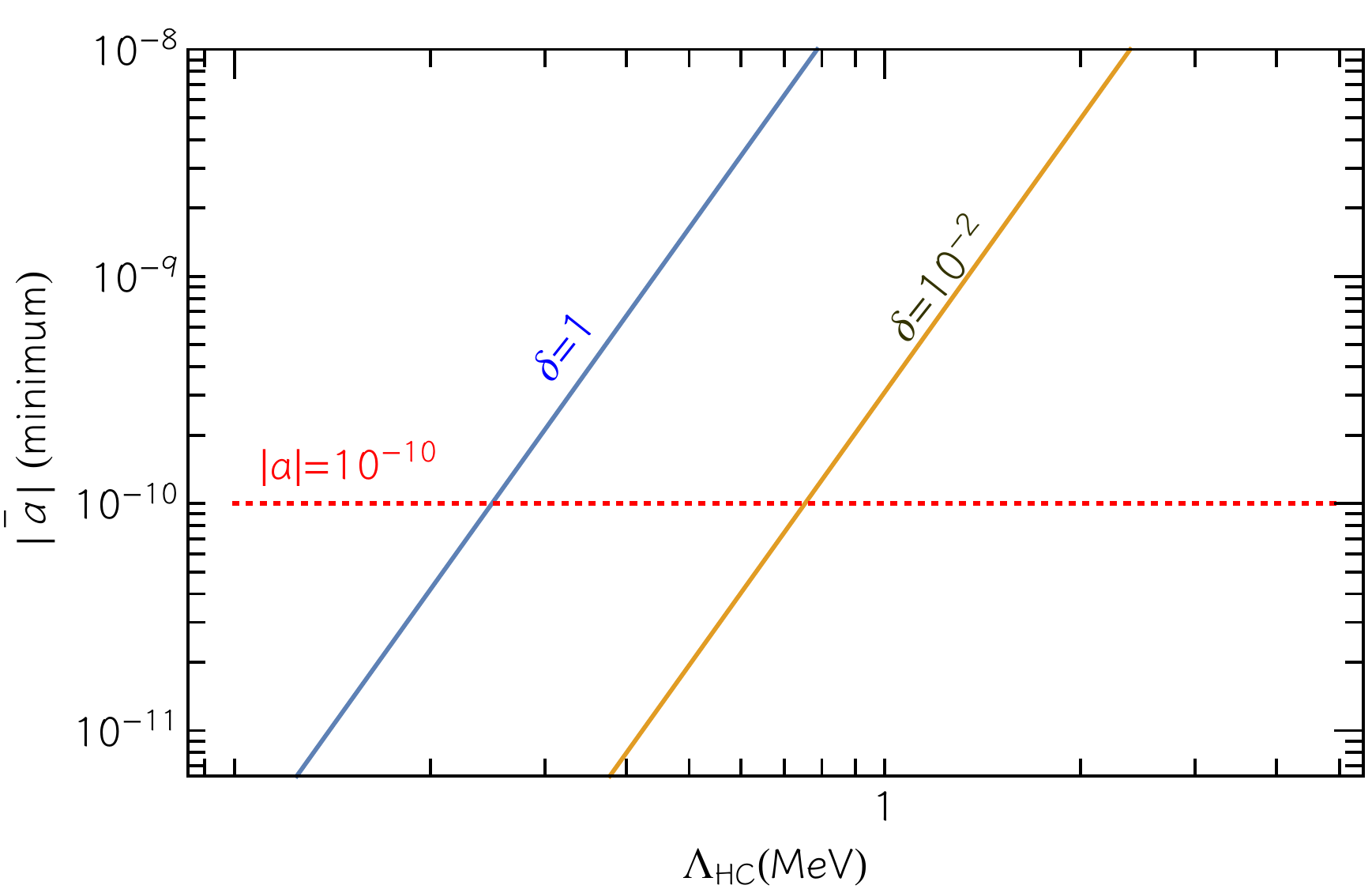}
  \vspace{-.1cm}
\caption{\label{fig:self}\em
Minimum of the potential $V(a)$ as a function of the new scale $\Lambda_{HC}$. }
\label{Minimum}
\end{figure}

We see that in the limit $\Lambda_{HC}\ll\Lambda_{\text{QCD}}$, the effective \textit{CP}-violating phase behaves as $\theta_{\text{eff}}\propto\frac{\Lambda_{HC}^4}{\Lambda_{QCD}^4}$, as one would naively expect. If we assume that the parameter $\delta$ is not unnaturally small, $\delta \gtrapprox O(10^{-2})$, we find that in order to not spoil the solution to the strong \textit{CP} problem the new scale needs to be
\begin{equation}
\Lambda_{HC} \lesssim 1\text{ MeV}	\,.
\end{equation}
Figure \ref{Minimum} shows the minimum of the potential as a function of $\Lambda_{HC}$ for different values of the phase $\delta$. When the HC potential is switched off, the minimum of the potential lies at $a=0$, as the solution to the strong \textit{CP} problem requires. Then, when the HC potential is switched on, the minimum is shifted proportionally to $\sim (\Lambda_{HC}/\Lambda_{QCD})^4$. As already stated, requiring $|a|\lesssim 10^{-10}$ implies an upper bound $\Lambda_{HC}\lesssim 1$ MeV for $\delta \gtrapprox O(10^{-2})$. Of course, the smaller (and more fine tuned) the shift $\delta$, the larger the upper limit (see the yellow curve in Fig.\ref{Minimum}). 

We stress here that the value of the new gauge coupling at high energies is of course also not arbitrary. A reference range for the example discussed in the following [a dark $G_{HC}=SU(3)$ with fermions much heavier than $\Lambda_{HC}$] can be taken to be 
\begin{equation}
	0.013 \lessapprox \alpha_{HC}(\Lambda_{UV} \sim 10^{11} \text{GeV}) \lessapprox 0.02\,,
\end{equation}
where $\Lambda_{UV}$, imagined to be set by inflationary physics, has been taken larger than the typical values of $f_a$ considered in this work. Of course, the appropriate range for $\alpha_{HC}(\Lambda_{UV})$ depends on the gauge group $G_{HC}$ and the matter content of the HC sector.

\section{Axion Dark Matter Abundance}
One of the most attractive features of the PQ solution of the strong \textit{CP} problem is that it provides a new dark matter candidate, the axion, to which great experimental and theoretical effort has been devoted \cite{Armengaud:2014gea, Bauer:2017ris, Caputo:2018ljp, Sikivie:2013laa, Sikivie:1983ip, Safdi:2018oeu, Kahn:2016aff, Hook:2018iia, Graham:2011qk, Budker:2013hfa, Sigl:2017sew, Raffelt:2006cw, Cameron:1993mr, Kelley:2017vaa, Shokair:2014rna, Brubaker:2016ktl, Brubaker:2017rna, Majorovits:2016yvk, Flambaum:2018ssk, Stadnik:2017hpa, Irastorza:2013dav, Irastorza:2018dyq, Caputo:2018vmy}. In typical models, when DWs are short lived the three nonthermal contributions (the decay of DW, coherent oscillations and cosmic string radiation) are comparable \cite{Kawasaki:2014sqa}. However, if DWs survive for a finite amount of time their contribution is enhanced. This is because while string decay and coherent oscillations occur when the Hubble parameter becomes comparable to the axion mass, $H\sim 3m_a$, DWs will still be topologically protected and will continue to evolve with cosmic expansion. When the axion starts to \textit{feel} the IIE (around $H\sim  H_{\text{decay}}$), the walls will decay into nonrelativistic axions. If $H_{\text{decay}}\ll m_a$, the contribution from domain wall decay dominates over the other contributions and, by imposing $\Omega_{\text{DW}}h^2\leq 0.12$, we can therefore constrain $H_{\text{decay}}$. This will automatically constrain $\Lambda_{HC}$ and $f_a$ due to Eq.(\ref{Hdecay}).

Assuming the so-called exact scaling regime we can write the relic abundance of axions today due to wall decay as
\begin{equation}
	\rho_a(t_0)=\Big(\frac{a(t_{\text{decay}})}{a(t_0)}\Big)^3\rho_a(t_{\text{decay}})\,,
\end{equation}
where 
\begin{equation}
	\rho_a(t_{\text{decay}})=\frac{\sigma}{t_{\text{decay}}}=\sigma H_{\text{decay}}\,,
\end{equation}
is the energy of the domain walls, then converted into axion, at the time of their decay. The relic density of cold axions from DWs is given by
\begin{equation}
	\Omega_{a,w}h^2=\frac{\rho_a(t_0)h^2}{\rho_c}=\Big(\frac{a(t_{\text{decay}})}{a(t_0)}\Big)^3\rho_a(t_{\text{decay}})h^2/\rho_c\,.
\end{equation}
It is useful to write the ratio
\begin{equation}
\frac{a(t_{\text{decay}})}{a(t_0)}=\frac{a(t_{\text{decay}})}{a(t_{\text{eq}})}\frac{a(t_{\text{eq}})}{a(t_0)}\,,
\end{equation}
and then use the relations 
\begin{equation}\begin{split}
&\frac{a(t_{\text{decay}})}{a(t_{\text{eq}})}=\Big(H(t_{\text{eq}})^2/2H(t_{\text{dec}})^2)\Big)^{1/4}\,,\\&
\frac{a(t_{\text{eq}})}{a(t_0)}=4.15\times 10^{-5}(\Omega_{\text{CDM}}h^2)^{-1}\,.
\end{split}\end{equation}
The Hubble parameter at matter-radiation equality and the critical density are given by
\begin{equation}\begin{split}
&H(t_{eq})=1.13\times 10^{-26}(\Omega_{CDM}h^2)^2\text{eV}\,,\\&
\rho_c=1.053\times 10^{-11}\Big(H/100\Big)^2\text{eV}^4\,.
\end{split}\end{equation}

\begin{figure}[!htb!]
\centering
\includegraphics[width=0.9\linewidth]{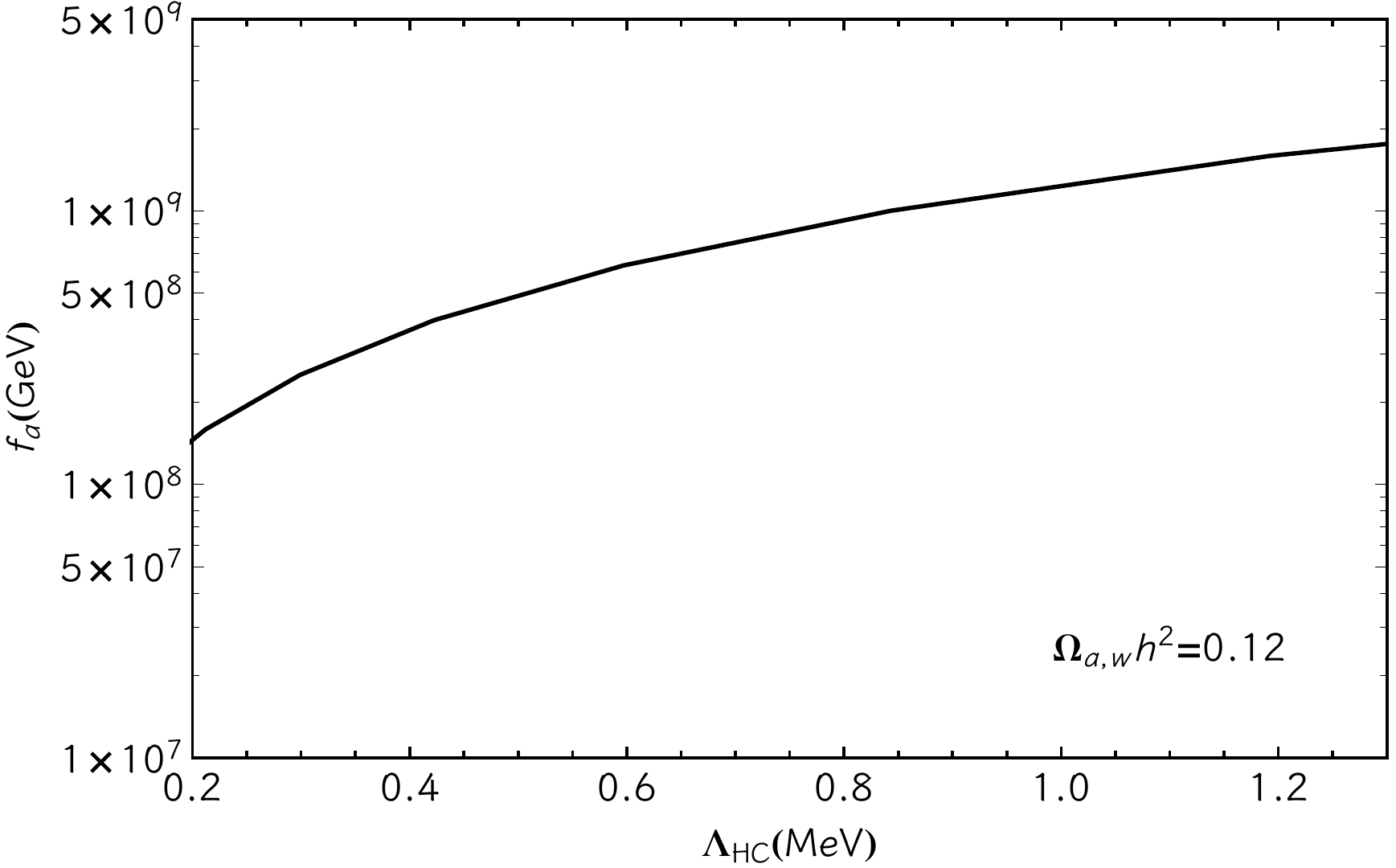}
  \vspace{-.1cm}
\caption{\label{fig:self}\em
Values of $\Lambda_{HC}$ and $f_a$ for which the axion abundance from domain wall decay matches the dark matter relic abundance today.}
\label{darkmatter}
\end{figure}

Finally, we find
\begin{eqnarray*}\label{wallaxion}
	\Omega_{a,w} =0.12\times\Big(\frac{f_a}{1.8\times10^9\text{ GeV}}\Big)^{3/2}\Big(\frac{\Lambda_{HC}}{\text{MeV}}\Big)^{-2}
\end{eqnarray*}
where we used
\begin{equation}
	m_a\sim 5.7\Big(\frac{10^9\text{ GeV}}{f_a}\Big)\text{ meV}\,.
\end{equation}
In Fig.\ref{darkmatter} we show the relation between the scale $\Lambda_{HC}$ and the decay constant $f_a$ for $\Omega_{a,w}h^2=0.12$ in a generic model with $N_{HC}=2$.
 Axions from coherent oscillations and radiated by strings can be estimated as \cite{Kawasaki:2014sqa}
\begin{equation}
\begin{split}
&\Omega_{coh} h^2\sim  0.0006\times \left(\frac{f_a}{1.8\times 10^{9}\text{ GeV}}\right)^{7/6}\,,   \\&
\Omega_{string} h^2\sim  0.001\times N_{DW}^2\left(\frac{f_a}{1.8\times 10^9\text{ GeV}}\right)^{7/6}\,.
\end{split}
\end{equation}
One can easily see that the axions coming from the decay of the walls will dominate the relic density today unless we have a large value of $N_{\text{DW}}^2$. In this case axions radiated from strings become important. \footnote{We warn the reader that the subject of axions from topological defects is of course a technically complicated one, to which a lot of effort has been dedicated (for an incomplete list of works, see Refs. \cite{Gorghetto:2018myk, Klaer:2017qhr,Hiramatsu:2010yu, Battye:1993jv, Fleury:2015aca, Fleury:2016xrz, Hagmann:1990mj, Yamaguchi:1998iv}); therefore, it follows that any estimate has to be taken with a grain of salt.}

To obtain a lower bound for $\Lambda_{HC}$ from DM abundance one has to distinguish between different axion models. This is because different lower bounds or constraints for the axion decay constant $f_a$ will pose different lower bounds for $\Lambda_{HC}$. 
If one considers the Dine-Fischler-Srednicki-Zhitnitsky (DFSZ) model, processes involving the axion coupling to electrons contribute to fast stellar cooling. From the constraint $g_{ae}\leq 2.6\times 10^{-13}$ \cite{Irastorza:2018dyq}, one gets 
\begin{equation}
f_a\geq 6.5\sin^2\beta \times 10^8\text{ GeV\,,}
\end{equation}
 where $\tan\beta=v_u/v_d$ is the ratio of up-type and down-type Higgs doublets in the DFSZ model.
For hadronic KSVZ models one has the constraints coming from SN1987A, where processes like $N\,N\rightarrow N\,N\,a$ generate a more efficient energy-loss channel, resulting in a reduced neutrino burst duration. This constrains the axion decay constant to be $f_a\geq 4\times 10^8$ GeV \cite{Raffelt:2006cw}.  For the sake of clarity, in the rest of the paper we will consider a general KSVZ-like model with $N_{\text{QCD}}\neq 1$ and heavy, vector-like quarks which we denote as $Q$.
\section{On the HC sector }\label{HC_section}
The dark HC sector can have a rich structure and many groups can lead to the desired interference effect, making this solution quite generic (up to simple model building). Here, for definiteness, we limit our analysis to the case of a dark $SU(N)_{HC}$ with a dark fermion $F$,  also charged under PQ symmetry. Guided by minimality, we assume that the fermions $F$ with HC charge receive their masses from the same scalar that breaks PQ symmetry, $\Phi$, which also gives masses to the colored fermions $Q$ of a generic KSVZ model. We also stress that the situation is quite different from other models with a non-Abelian dark sector where the new degrees of freedom also carry other SM charges (see, for example, Refs. \cite{Contino:2018crt,Mitridate:2017oky, Antipin:2015xia, Cirelli:2005uq,Kribs:2016cew, DeLuca:2018mzn}). For us, all the particles of the dark HC sector are singlets under $SU(3)_C\times SU(2)_L\times U(1)_Y$. Moreover, we expect the mass of the dark sector fermion $F$ to be
\begin{equation}
	m_F \sim Y_1 v\,,
\end{equation}  
where $Y_1$ is a Yukawa coupling and $v=N_{\text{QCD}}f_a$ is the scale of the spontaneous breaking of PQ symmetry. Requiring that the Yukawas not be too small (in order to not introduce again a small parameter as the $\theta$ term) it is easy to see that we will always be in the situation where the mass of the fermions is much larger than the confinement scale, 
\begin{equation}
	m_F\gg \Lambda_{HC} \sim 1\text{ MeV}\,.
\end{equation}
In this case one can have stable, heavy bound states composed of fermions of the HC sector. These bound states contribute to the energy density of the Universe and (depending on its mass and time of decoupling) can behave as cold dark matter. In addition, as we show in the next section, we can have stable glueballs that might be cosmologically dangerous.

\begin{figure}[!htb!]
\centering
\includegraphics[width=1.0\linewidth]{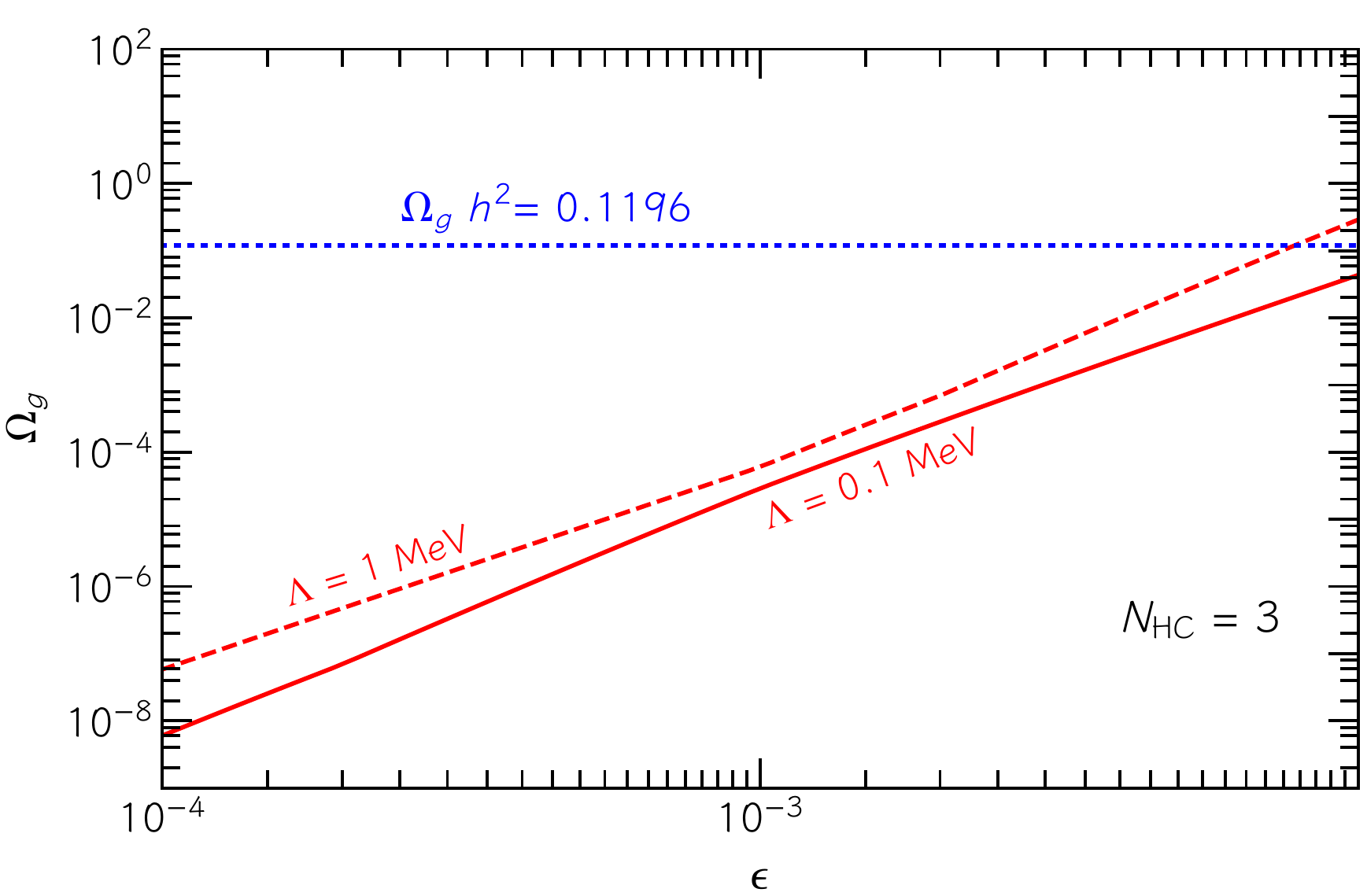}
  \vspace{-.1cm}
\caption{\label{relic_glueball}\em
Relic density of glueballs for $N_{HC}=3$ and two different confinement scales, $\Lambda= 0.1, 1 $ MeV.
}
\end{figure}

\subsection{HC glueballs and glueballinos}
 The details of the cosmology of the model depend on the representation we choose for the fermions. If the new fermion is in the fundamental representation of $SU(N)_{HC}$, for example, the spectrum of the theory will contain HC glueballs and HC hadrons (hyperbaryons) composed of $N$ heavy fermions $F$. A glueball in the dark sector is a bound state of HC gluons with a mass of the order of the confinement scale $\Lambda_{HC}$ and strongly self-interacting. If glueballs are stable, their relic density can be computed in terms of only two parameters: $\Lambda_{HC}$ and $\epsilon=T_{HC}/T$. $T_{HC}$ is the temperature of the dark sector, which does not necessarily coincide with the SM temperature $T$. The relic density of glueballs is given by \cite{Boddy:2014yra}
	\begin{equation}
	\Omega_gh^2\sim \frac{s_0}{\rho_{c,0}/h^2}\frac{2(N_{HC}^2-1)}{g_{\star\,S}(T_f)}\epsilon^3\Lambda_{HC}\,.
	\end{equation}
We can easily see that this relic density becomes larger than the DM relic density unless $\epsilon$ is small. This is known as the dark glueball problem \cite{Halverson:2016nfq,Halverson:2018olu}. In Fig.\ref{relic_glueball} we  show the glueball relic density for two different values of the confinement scale in the range of interest. We see that the temperature ratio has to be $\epsilon \lessapprox 10^{-3}-10^{-2}$ in order to not overclose the Universe.

We notice that if the dark fermion is in the fundamental representation of $SU(N)_{HC}$, it will be $N_{HC}=1$. While this of course leads to the IIE and the resolution of the domain wall problem, it could seem as arbitrary as taking $N_{\text{QCD}}=1$ from the beginning. As an example we explore a more general situation in which the fermion is in the adjoint representation, $F\sim(\text{Adj}, 1,1,0)$. In such a case the spectrum will be composed of HC glueballs and glueballinos (as in Ref. \cite{Boddy:2014yra}) and the HC anomaly coefficient will be $N_{HC}=N$. Glueballinos are composite states of an HC gluon and a fermion in the adjoint representation, $(gF)$, which are protected from decay by gauge and Lorentz symmetry. In both, the fundamental and adjoint cases, the hadron-like states are expected to be much heavier than the HC scale $M_{hadron}\sim m_F$ and self-interact strongly by glueball exchange of masses $m\sim \Lambda_{HC}$. The relic density today of glueballinos or HC hadrons can be estimated in a canonical way, as it is done for other WIMP candidates (taking into account only \textit{S}-wave annihilation)\cite{Boddy:2014yra,Feng:2008mu}:
	\begin{equation}
	\Omega_Fh^2\sim\frac{s_0}{\rho_{c,0}}\frac{25\epsilon}{(g_{\star\,s}/\sqrt{g_{\star}^{tot}})M_P\vev{\sigma v}}\,.
	\label{glueballinorelic}
	\end{equation}
	where we also take $\vev{\sigma v}\sim \alpha_{HC}^2(m_F)/m_F^2$. This expression neglects the effect of reannihilation considered in Ref. \cite{Contino:2018crt}. A more sophisticated expression should take this effect into account, investigating both deexcitation and dissociation processes. However, in the present setup glueballs do not decay and therefore we expect reannihilation to be inefficient. Consequently, we will neglect its effects. 
	
	Concerning the glueballino relic density, we note that, as usually happens for stable relics that were in thermal equilibrium, they can overclose the Universe if their mass is larger than $\mathcal{O}(100)$ TeV \cite{Griest:1989wd}. In addition, our scenario also suffers from the glueball problem and, therefore, one needs $\epsilon\ll 1$. In such a case, when $m_F\ll 100$ TeV and $\epsilon\ll 1$, the relic density of dark matter will be QCD axion dominated. In a following section we will identify the regions of parameter space where dark matter is composed of axions and those where dark matter is made of glueballs and glueballinos. 
	
	Finally, let us note that the suppression of the dark sector temperature respect to the Standard Model also allows to evade bounds from the number of relativistic degrees of freedom $N_{eff}$. Results from the Planck Collaboration give $N_{eff} = 2.99 \pm 0.17$ \cite{Aghanim:2018eyx}. We then have to worry about whether our framework implies nonstandard values of $N_{eff}$.  The temperature of the Standard Model when the dark sector confines is given by
	\begin{equation}
		T_{conf}=100 \, \text{MeV} \Big(\frac{T_{HC}}{\text{MeV}}\Big)\Big(\frac{0.01}{\epsilon}\Big)\,.
	\end{equation}
	Therefore, for $\epsilon \lessapprox 10^{-2}$ there are no relativistic, massless species to act as the hidden-sector bath during BBN because HC confinement occurs well before BBN and structure formation. 	

\begin{figure}[!htb!]
\centering
\includegraphics[width=1.05
\linewidth]{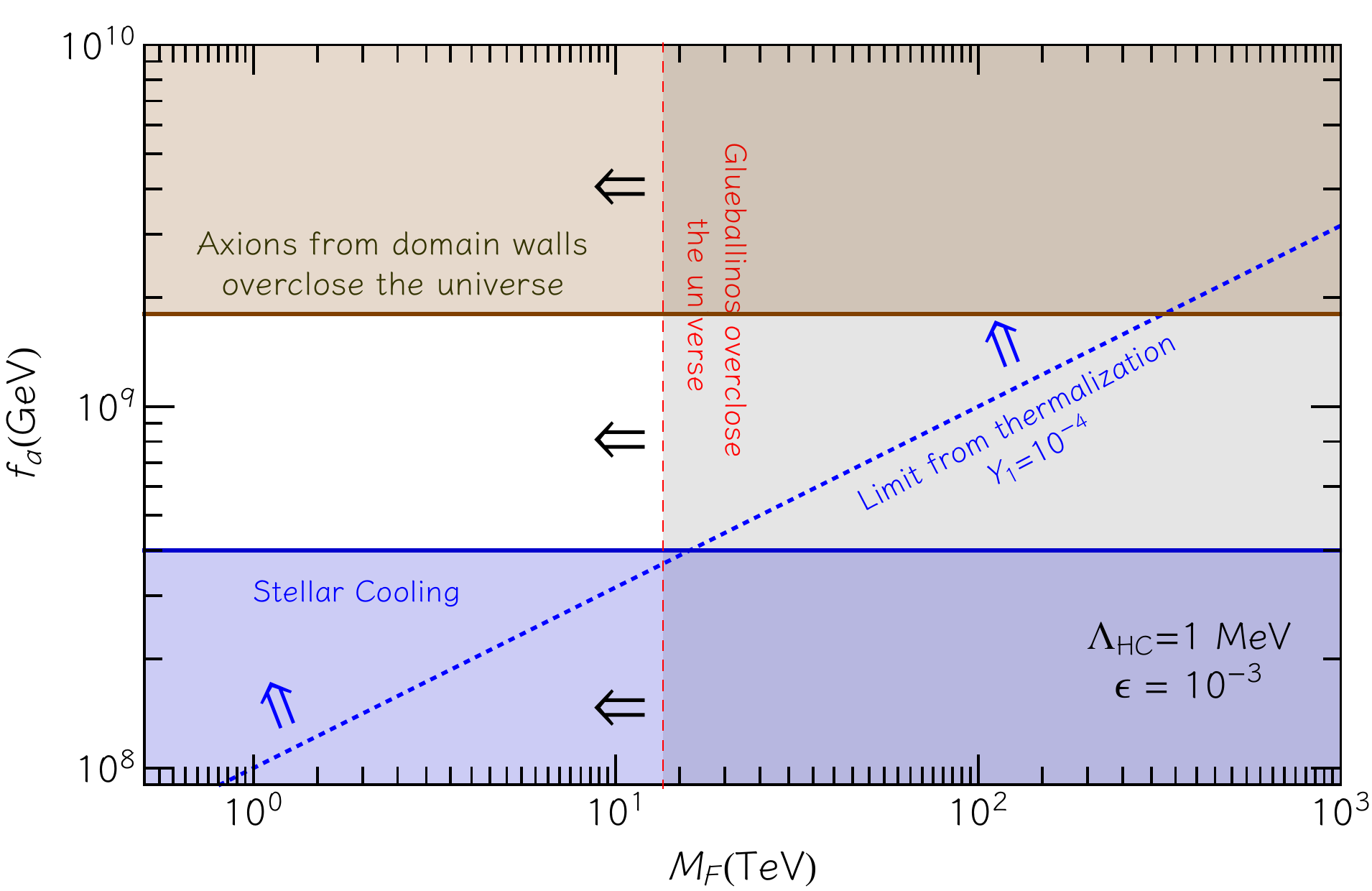}
  \vspace{-.1cm}
\caption{\label{resume_plot}\em
Here we show the constraints on the parameters $f_a$ and $M_F$ for fixed values of $\Lambda_{HC}$ and $\epsilon$. The bound from stellar cooling \cite{Raffelt:2006cw} is of course independent of the latter (and also of $M_F$). For large values of the decay constant, $f_a \gtrapprox 2\cdot 10^{9}\text{ GeV}$, axions are overproduced and overclose the Universe. Similarly for large values of the dark fermion mass, $M_F \gtrapprox 13.5 \text{ TeV}$, glueballinos overclose the Universe.
}
\end{figure}

\subsection{Thermal equilibrium between the HC sector and the Standard Model}
In the previous section we introduced a new parameter $\epsilon=T_{HC}/T$. The glueball and/or glueballino relic abundance in the considered model turn out to be an important problem unless $\epsilon$ is small. This means that the HC sector needs to have a smaller temperature than the SM thermal bath. We make the assumption that the two sectors start with different equilibrium temperatures\footnote{This might occur due to their coupling with the inflaton sector.}. Still, we then have to make sure that the two sectors never enter thermal equilibrium. The strongest constraint on the parameters of the theory comes from the scattering process
\begin{equation}
	g_{HC} + F \rightleftarrows a + F
\end{equation}
mediated by a heavy fermion $F$. The rate for this process at high temperatures is given by
\begin{equation}
	\Gamma=T^3\frac{\alpha_{HC}}{T^2}\Big(\frac{m_F}{f_a}\Big)^2\,,
\end{equation}
which has to be compared with the Hubble rate
\begin{equation}
	\Gamma_H= H=\sqrt{\frac{4\pi^3G g_*}{45}}T^2=\sqrt{\frac{g^*}{100}}\frac{T^2}{7.36\cdot 10^{17} \text{ GeV}}\,.
\end{equation}
In the above equation $G$ is Newton's constant, $T$ is the temperature of the Standard Model and $g^*$ is the number of relativistic degrees of freedom, which is given by
\begin{equation}
	g^*(t)=g^*_{SM}(t)+g^*_{HC}(t)\Big(\frac{T_{HC}(t)}{T(t)}\Big)^4 \sim g^*_{SM}(t)\,.
\end{equation}
The comparison can be made for $T_{HC}=m_F=\epsilon T$, that is to say when $T=m_F/\epsilon$. Below this temperature the number density of heavy fermions $F$ is rapidly suppressed by the Boltzmann factor. Expressing $m_F=Y_1f_a$, we find that in order to not enter thermal equilibrium
\begin{equation}
	Y_1 \lessapprox 10^{-5} \Big(\frac{0.01}{\alpha_{HC}}\Big)\Big(\frac{0.01}{\epsilon}\Big)\Big(\frac{f_a}{10^9\text{ GeV}}\Big)\,,
	\label{constraint}
\end{equation} 
which also implies 
\begin{equation}
	m_F\lessapprox 10\text{ TeV}  \Big(\frac{0.01}{\alpha_{HC}}\Big)\Big(\frac{0.01}{\epsilon}\Big)\Big(\frac{f_a}{10^9\text{ GeV}}\Big)^2\,.
\end{equation}
Notice that $\alpha_{HC}\equiv\alpha_{HC}(m_F)$ is the HC gauge coupling at the $F$ mass scale. We also checked other possible processes such as the scattering of colored and dark fermions mediated by the axion (or by the heavy scalar), scattering of dark and visible gluons, scattering of Higgses and heavy scalars, etc. For all of these we still find that Eq.(\ref{constraint}) gives the strongest constraint on $Y_1$. 

\section{Dark matter candidates}
As we have seen, the solution of the domain wall problem actually introduces new dark matter candidates, namely, glueballs and glueballinos. It is therefore interesting to explore the range of parameters to see where the axions are (or are not) the dominant dark matter candidates. Indeed, even if we add new parameters, most of them are highly constrained: the new confinement scale has been found to be $\Lambda_{HC} \sim 0.1 - 1$ MeV, the ratio of the temperatures of the two sectors has to be $\epsilon \lessapprox 10^{-2}-10^{-3}$ in order to not overproduce glueballs, and the Yukawa of the HC fermion is $Y_1 \lessapprox 10^{-5} \Big(\frac{0.01}{\alpha_{HC}}\Big)\Big(\frac{0.01}{\epsilon}\Big)\Big(\frac{f_a}{10^9\text{ GeV}}\Big)$ in order not to spoil thermal (non)equilibrium. To explore the parameter space we fix $\Lambda_{HC}=1 \text{ MeV}$ and $\epsilon=10^{-3}$. The latter value makes glueball relic density almost negligible. On the one hand, for large values of the axion decay constant
\begin{equation}
f_a \gtrapprox 1.8\cdot 10^{9}\Big(\frac{\Lambda_{HC}}{1\text{ MeV}}\Big)^{4/3} \text{ GeV}\,,
\end{equation}
axions from domain wall decay overclose the Universe [see Eq.(\ref{wallaxion})]. On the other hand, SN1987A gives a lower limit of $f_a \gtrapprox 4\cdot 10^{8}$ GeV \cite{Raffelt:2006cw}. It follows that the value of $f_a$ is restricted to be in a small window (which may be testable in the near future), which is thinner for smaller values of the confinement scale (which correspond to larger values of the phase $\delta$). In addition, large values for the dark fermions masses ($M_F \gtrapprox 13.5\,$ TeV) are also excluded (see Fig. \ref{resume_plot}) in order to not overclose the Universe. In the remaining parameter space one can arrange $M_F$ and $f_a$ to have dark matter mostly in axions or glueballinos. Notice also that the allowed region is always far from the thermalization bound (dotted blue line in Fig.\ref{resume_plot}), which ensures that the treatment is consistent. As a simple rule of thumb, regions where $f_a$ is large and $M_F$ is small are axion dominated, while regions with large $M_F$ and small $f_a$ are glueballino dominated. It is also interesting to notice that the solution of the domain wall problem and the consequent value of the confinement scale $\Lambda_{HC}$ naturally led us to an interesting region for glueballs and glueballinos, which was extensively studied in Ref. \cite{Boddy:2014yra} with the goal of introducing self-interacting dark matter as  a solution to the small-scale formation woes in astrophysics \cite{Liu:2019bqw, Spergel:1999mh}. As shown in Ref. \cite{Boddy:2014yra}, it is also possible to arrange the parameters of the model (just by slightly increasing $\epsilon$) to have most of the dark matter in the form of glueballs instead of glueballinos. Ultimately, what matters is the fact that the dark sector does not communicate with the visible one and the entropy in the dark sector is separately conserved. Then one can play to have more dark matter in one dark species than another, but this does not alter the overall picture.  
 
\section{Gravitational waves from the dark}
Violent phenomena in the early Universe can lead to large anisotropic fluctuations in the energy-momentum tensor and therefore gravitational waves (GWs). An example of such phenomena are first-order phase transitions \cite{Hogan:1986qda, Witten:1984rs, Turner:1989vc}. Within the Standard Model there are at least two phase transitions: one associated with the breaking of the electroweak symmetry and one with the breaking of chiral symmetry. Both of them are known to proceed through a smooth crossover, that is to say, they are not first order and cannot be sources of GWs. One may wonder if in our model the HC sector [$SU(N)_{HC}$]  phase transition can lead to a GW signal in the frequency range of future detectors \cite{Schwaller:2015tja}. The situation under consideration, $m_F \gg \Lambda_{HC}$, is analogous to a pure Yang-Mills theory, which is well known to exhibit a strong first-order phase transition \cite{Svetitsky:1982gs}. However, a GW signal is proportional to the energy density of the dark sector, which is proportional to the fourth power of the temperature $T_h$. As we have seen, the latter is smaller than the temperature of the visible world and the GW signal will be consequently suppressed. Using the same input parameters and the results of Ref. \cite{Schwaller:2015tja}, and properly taking into account the suppression of the energy density due to the low HC sector temperature, we find that the signal frequency will fall within the range of frequency of SKA \cite{WinNT} (thanks to the difference in temperatures between the dark and visible sectors), 
\begin{equation}
	f_{peak}\approx 3.33\cdot 10^{-9}\,Hz\Big(\frac{\Lambda_{HC}}{1 \text{ MeV}}\Big)\Big(\frac{10^{-2}}{\epsilon}\Big)
\end{equation}
but just outside its sensitivity for the considered values of $\epsilon$. Nevertheless, this is an interesting situation that deserves further study and will be presented elsewhere.\\
Finally, we also checked the possible GW production from domain wall decays into gravitons \cite{Saikawa:2017hiv}, which unfortunately turns out to be subdominant and always negligible in the explored parameter space.

\section{Conclusions}
We have explored a scenario where the domain wall problem of axion models is naturally solved by introducing a new gauge group which confines at a scale $\Lambda_{HC}$ smaller than $\Lambda_{\text{QCD}}$. This scenario does not require extremely small parameters or inverse phase transitions, and turns out to be quite general. Constraints from the domain wall energy density and the requirement of not spoiling the solution to the strong \textit{CP} problem fix the HC scale to be $\sim$ MeV. 

Interestingly enough, this particular scale and the mechanism itself have been shown to have a number of nontrivial consequences. For a simple and minimal model in which $HC=SU(N)_{HC}$, we found that having $\Lambda_{HC}\sim 1$ MeV naturally provides new self-interacting dark matter candidates in the form of glueballs or glueballinos (or both) with masses below $\sim 10$ TeV. Moreover, the HC sector turns out to have a different temperature ($T_h$) from the visible one, making it possible for $\Lambda_{HC}\sim 1$ MeV to produce gravitational waves in the frequency range of SKA due to the strong first-order phase transition of $SU(N)_{HC}$. Finally, it is noteworthy that axions are mainly produced from domain wall decay and, unless one assumes unnaturally small values for the phase $\delta$, the decay constant $f_a$ is constrained to be $4\cdot 10^8\text{ GeV} \leq f_a \lessapprox 2\cdot 10^{9}\text{ GeV}$. This corresponds to an axion mass range of $1-15\text{ meV}$. It is remarkable that this lies close to the expected sensitivity of \textit{axion antennas} \cite{Lawson:2019brd}, dielectric haloscopes \cite{Millar:2016cjp}, or ARIADNE \cite{Arvanitaki:2014dfa}. Axion experiments will therefore be able to test this scenario in the near future.

\section{acknowledgments}
	We warmly thank Andrea Mitridate and Masaki Yamada for useful discussions and for reading the manuscript. We also thank Ciaran O'Hare, Javier Redondo, Ken'ichi Saikawa, Marco Taoso and Pablo Villanueva-Domingo for enlightening discussions.\\
	M.R. is supported by the Spanish grants SEV-2014-0398 and FPA2017-85216-P (AEI/FEDER, UE), PROMETEO/2018/165 (Generalitat Valenciana), the Spanish Red Consolider MultiDark FPA2017-90566-REDC and FPU grant FPU16/01907.
	A.C. is supported by grants FPA2014-57816-P, PROMETEOII/2014/050 and SEV-2014-0398,  as well as by  the EU projects H2020-MSCA-RISE-2015 and H2020-MSCA-ITN-2015//674896-ELUSIVES.

%\bibliographystyle{utphys}
%\bibliography{newrefs,newrefs_axion} 

\providecommand{\href}[2]{#2}\begingroup\raggedright\endgroup

\end{document}